\documentstyle[sprocl,epsfig]{article}
\input{epsf.tex}

\def\Journal#1#2#3#4{{#1} {\bf #2}, #3 (#4)}

\def\NPB{{\em Nucl. Phys.} B}
\def\PLB{{\em Phys. Lett.}  B}
\def\PRL{\em Phys. Rev. Lett.}

\def\be{\begin{equation}}
\def\ee{\end{equation}}
\def\bea{\begin{eqnarray}}
\def\eea{\end{eqnarray}}
  \newcommand{\epsfaxhax}[2]{
          \centerline{
            \hspace{-15pt}
            \epsfxsize=160pt
            {\epsfbox{#1}}
            \hspace{-7pt}
            \epsfxsize=160pt
            {\epsfbox{#2}}}
  }

\begin{document}

\title{ MEASURING DIQUARK CONDENSATION IN LATTICE SIMULATIONS OF DENSE MATTER}

\author{Simon HANDS and Susan MORRISON}

\address{Department of Physics, University of Wales Swansea,\\ Singleton Park, 
Swansea SA2 8PP, UK.\\E-mail: s.hands,s.morrison@swansea.ac.uk} 


\maketitle\abstracts{ We discuss general aspects of the possibility of Bose
condensation of diquark pairs in systems of dense matter. Lattice 
field theory simulations are presented for model four-fermion theories
which are expected to manifest the phenomenon, and results from 
measurements of both diquark two-point and one-point functions presented.
Whilst inital results are promising, there remain systematic effects needing
to be
understood.}


There has been much recent interest in the possibility of Bose condensates
formed from quark pairs close to the Fermi surface in dense matter; such 
condensates, which carry non-zero baryon number, can lead to exotic scenarios
for dynamical breaking of the color gauge symmetry \cite{Krishna}. So far
most work in this direction has relied on effective field theory descriptions
of the strong interaction. For any problem invloving the QCD ground state 
it would be nice to have a calculation which in principle involves
no uncontrolled approximations;
this is therefore a natural area for the application of the methods of
lattice gauge  theory.
In this talk we describe the first steps we have made in this direction;
our initial results have already appeared \cite{HM}.

When thinking about possible diquark condensates $\langle
qq\rangle\not=0$ which might form there are several issues to consider.
Firstly, is the condensate wavefunction gauge invariant? In the QCD
scenarios the condensate is formed
from a $\bf{3}\otimes\bf{3}$ of the color gauge
group, which 
necessarily 
breaks the local symmetry, leading to a dynamical Higgs mechanism making some
or all of the gluons massive. This is the phenomenon of {\sl color
superconductivity\/}. In other models, relevant to this discussion, 
another possibility
is that the condensate wavefunction is gauge invariant; in this case 
condensate formation breaks only global symmetries, and 
yields a {\sl superfluid\/} state.
Next, 
is the condensate necessarily a spacetime scalar? We normally expect 
condensates to share the spacetime 
quantum numbers of the perturbative vacuum, but
in systems of dense matter there is a preferred rest frame, making it possible
to consider rotationally non-invariant condensates \cite{Krishna}. Another
possibility which should not be excluded {\it a priori\/} is that the condensate
spontaneously breaks parity invariance \cite{HM}.
Finally, the most crucial aspect of the condensate wavefunction is that it 
respects the Pauli Exclusion Principle, ie. it must be antisymmetric under 
exchange of all possible quantum numbers between the quarks. It is this 
requirement which makes the ground state so exquisitely sensitive to the number
of light quark flavors present in QCD \cite{Krishna}; more generally it may 
also result in a sensitivity to which representation of the gauge group 
is carried by the quarks.


The most generic symmetry we expect to be be broken by a diquark condensate 
is the global U(1)$_V$ of baryon number:
\be
q\mapsto e^{i\alpha}q\;\;\;;\;\;\;\bar q\mapsto\bar q e^{-i\alpha}.
\label{eq:B}
\ee
One issue we should address is  that spontaneous breaking of a 
global vectorlike symmetry like this is
usually forbidden by the Vafa-Witten theorem \cite{VW}. It is convenient to 
classify the models in which diquark condensation might occur by the various 
escape clauses offered by the theorem.
The theorem does not apply if the path integral measure is not positive
definite. This is the case for QCD with chemical potential $\mu\not=0$, since
for this case the fermion determinant is complex. Unfortunately it is precisely
this feature which makes Monte Carlo simulation extremely difficult.
The theorem also fails if the theory includes a Yukawa coupling 
to a scalar degree of freedom. Two model field theories satisfying this
criterion are the Gross-Neveu (GN) model in 2+1 dimensions, and the Nambu --
Jona-Lasinio (NJL) 
model in 3+1 dimensions, and these will be the main focus of this
talk.
The final escape route is if baryon number is part of some larger 
non-vectorlike symmetry. An example is SU(2) lattice gauge theory, where
the global symmetry group enlarges from $\mbox{U(1)}_V\otimes\mbox{U(1)}_A$
to $\mbox{U(2)}$ due to the pseudoreal nature of the {\bf 2} representation.
Simulations of SU(2) lattice gauge theory with $\mu\not=0$ will be discussed
separately \cite{MH}.



The GN and NJL models, which are essentially identical apart from the number of 
spatial dimensions, are relativistic generalisations of the model originally 
considered in the BCS mechanism for superconductivity. In continuum notation
the Lagrangian may be written:
\be
{\cal L}=\bar\psi(\partial{\!\!\! /}\,+m)\psi
-g^2[(\bar\psi\psi)^2-(\bar\psi\gamma_5\vec\tau\psi)^2].
\label{eq:action}
\ee
The four-fermi terms can be replaced by Yukawa couplings to scalar and
pseudoscalar auxiliary fields.
The lattice transcription of (\ref{eq:action}) is discussed in detail elsewhere
\cite{HM}. In addition to the U(1) baryon number symmetry (\ref{eq:B}), 
the model has a $\mbox{SU(2)}_L\otimes\mbox{SU(2)}_R$ axial symmetry in the
chiral limit $m\to0$:
\be
\psi_L\mapsto U\psi_L\;,\;\bar\psi_L\mapsto\bar\psi_LU^\dagger\;\;;\;\;
\psi_R\mapsto V\psi_R\;,\;\bar\psi_R\mapsto\bar\psi_RV^\dagger,
\label{eq:axial}
\ee
with $U,V$ independent SU(2) matrices.

The most relevant features of the model for our purposes are that for strong 
coupling the $\mbox{SU(2)}_L\otimes\mbox{SU(2)}_R$ symmetry spontaneously
breaks to $\mbox{SU(2)}_V$ by formation of a chiral condensate
$\langle\bar\psi\psi\rangle$. The spectrum in the broken phase contains both
``baryons'', namely the elementary fermions which now have a dynamically
generated mass, and ``mesons'', namely
$\psi\bar\psi$ composites, which include 3 Goldstone pions. It turns out that 
for the 2+1 dimensional GN case the model has an interacting continuum limit 
at the critical coupling required for symmetry breaking. Crucially, 
the model can be formulated on a lattice and simulated for $\mu\not=0$
\cite{Karsch}; it is found that in the broken phase 
a first order chiral symmetry restoring
transition occurs for some critical $\mu_c$ \cite{SKK}.

We have considered \cite{HM} the possible formation of a diquark condensate with
wavefunction
\be
qq=\psi^{tr}{\cal C}\gamma_5\otimes\tau_2\otimes\tau_2\psi;
\label{eq:qq}
\ee
the operators in the tensor product denote that the diquark is
a spacetime scalar, and antisymmetric in both implicit (due to lattice
fermion doubling) and explicit flavor indices (note that ${\cal C}\gamma_5$
is also antisymmetric). Condensation of (\ref{eq:qq}) spontaneously
breaks baryon number (\ref{eq:B}) but not axial (\ref{eq:axial}) symmetry.
We have examined two possible signals for $\langle qq\rangle\not=0$.


\begin{figure}[t]
\epsfaxhax{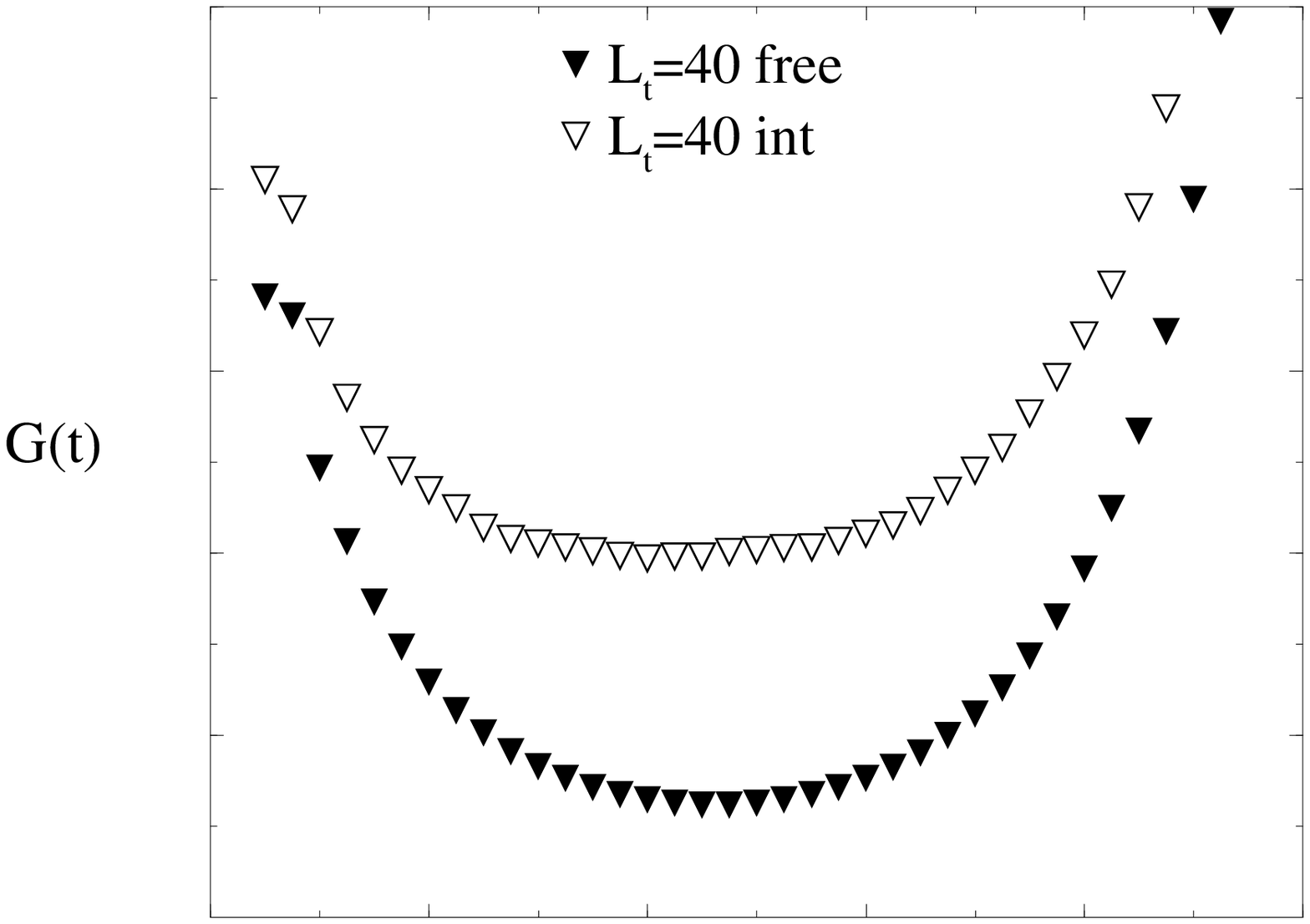}{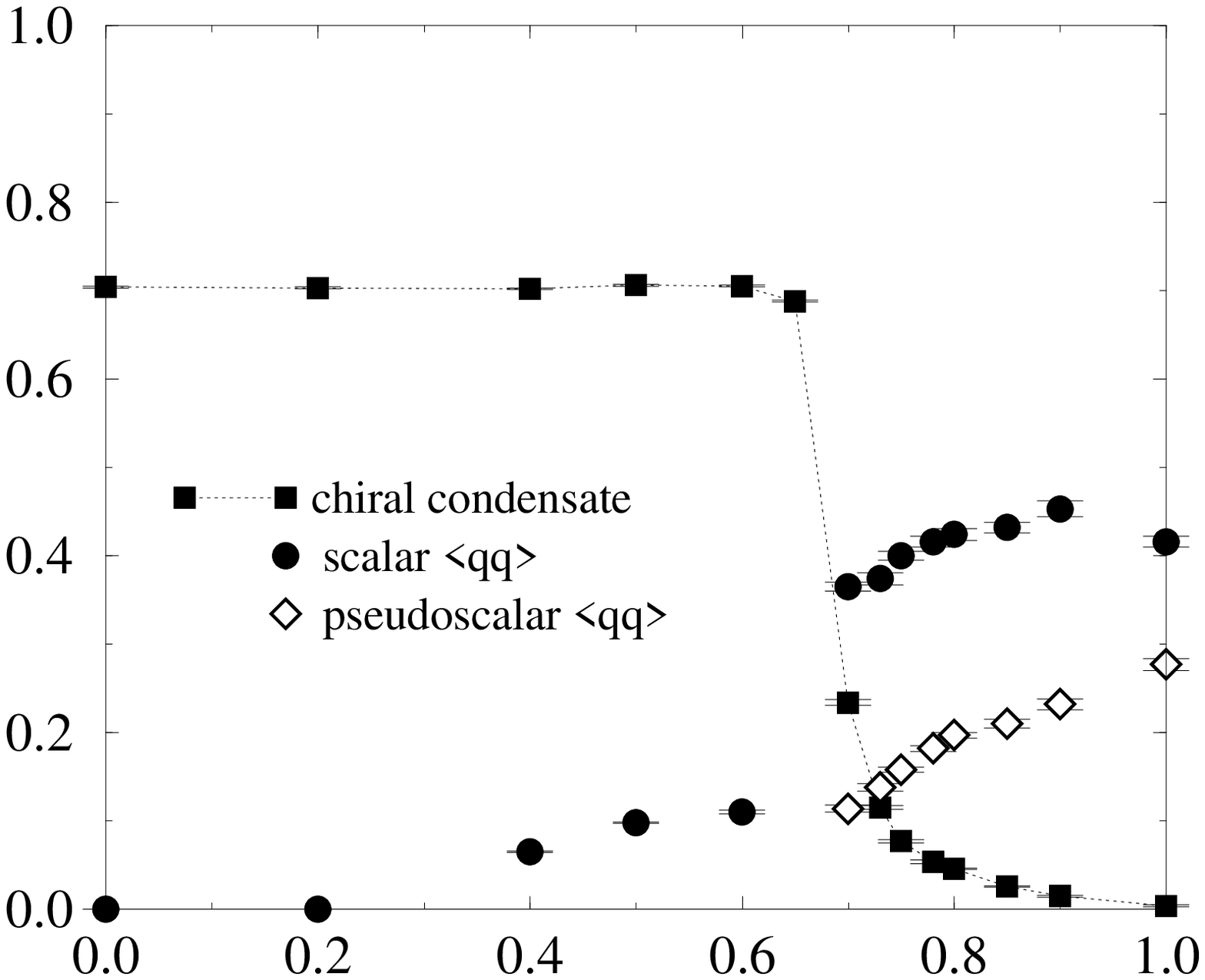}
\vskip -0.5cm
\caption{\hskip 4.1cm Figure 2:}
\vskip -0.5cm
\setcounter{figure}{2}
\end{figure}

For large spatial separation the diquark pair propagator is proportional to
the square of the condensate by the cluster property:
\be
G(x)=\langle qq(0)\bar q\bar q(x)\rangle=\langle qq(0)\bar q\bar q(x)\rangle_c
+\langle qq\rangle\langle\bar q\bar q\rangle\Rightarrow
\displaystyle\lim_{x\to\infty}G(x)=\vert\langle qq\rangle\vert^2.
\ee 
A non-zero condensate should therefore reveal itself as a plateau in the 
large-$t$ behaviour of the diquark timeslice propagator. In Fig. 1 we show
the behaviour of $G(t)$ from simulations of the GN model on a $16^2\times
40$ lattice, performed with a value $\mu=0.8>\mu_c$, 
so that chiral symmetry is restored. There is clear evidence for a stable
plateau, especially when compared to $G(t)$ for non-interacting
fermions, also shown with closed symbols. 
The square root of the plateau height is plotted, 
together with $\langle\bar\psi\psi\rangle$, as a function of $\mu$ in Fig. 2.
We observe a large increase in the $\langle qq\rangle$
signal going from the chirally 
broken phase into the symmetric phase, consistent
with the notion that the condensates ``compete''\cite{Krishna}. We also
observe a small but non-zero $\langle qq\rangle$ in the low density phase, 
and a parity violating pseudoscalar condensate in the high density phase, 
although probably both signals are finite volume artifacts.

Unfortunately the height of the plateau remains roughly constant as
the spatial volume of the lattice is increased, whereas naively we expect it
to be extensive. It is therefore not clear whether a true condensation is
occurring, or whether there are unexplained finite volume effects. 
A similar behaviour is seen in 3+1 dimensional simulations, 
suggesting that this is not a low dimensional artifact as originally thought
\cite{HM}.


To attempt to clarify these ambiguities we have now performed direct
measurements of $\langle qq\rangle$ in the simulations. This involves including
explicit diquark source terms in the action and using a 
{\sl Gor'kov\/} representation, ie:
\bea
S_{ferm} & = & \bar\psi
M\psi+j\psi^{tr}\tau_2\psi+\bar\jmath\bar\psi\tau_2\bar\psi^{tr} 
        =  (\bar\psi,\psi^{tr})\left(\matrix{\bar\jmath\tau_2&{1\over2}M\cr
      -{1\over2}M^{tr}&j\tau_2\cr}\right)\left(\matrix{\bar\psi^{tr}\cr\psi\cr}
\right)\nonumber\\
       &\equiv& \Psi^{tr}{\cal A}[j,\bar\jmath]\Psi;\\
Z[j,\bar\jmath] &= &
\bigl\langle\mbox{Pf}({\cal A}[j,\bar\jmath])\bigr\rangle.
\eea
The diquark condensate is now defined by 
\be
\langle qq\rangle={1\over V}{{\partial\ln Z}\over{\partial j}}\biggr\vert_
{j,\bar\jmath=0}=\displaystyle\lim_{j,\bar\jmath\to0}{1\over V}
\biggl\langle{1\over2}\mbox{tr}\left\{{\cal A}^{-1}\left(
\matrix{0&0\cr0&\tau_2\cr}\right)\right\}\biggr\rangle,
\ee
which is straightforward to implement. Our results are
``quenched'' in the sense that we have included $j\not=0$ in the measurement
routines but not as yet in the update algorithm. We show $\langle qq(j)\rangle$
for various $\mu$ for the GN model in Fig. 3 and for the
NJL model in Fig. 4. In the low density chirally broken phase the
signal extrapolates linearly to zero as $j\to0$; in the high density phase
the signal is larger and considerably less linear, although still vanishing
in the zero source limit. Preliminary indications are that the curvature 
of $\langle qq(j)\rangle$ is
still more marked in the 3+1 dimensional simulation. We have found evidence
for small 
finite volume effects, but the question of whether a diquark 
condensate forms, as revealed by a non-zero intercept in the
thermodynamic limit, is still open.

\begin{figure}[t]
\epsfaxhax{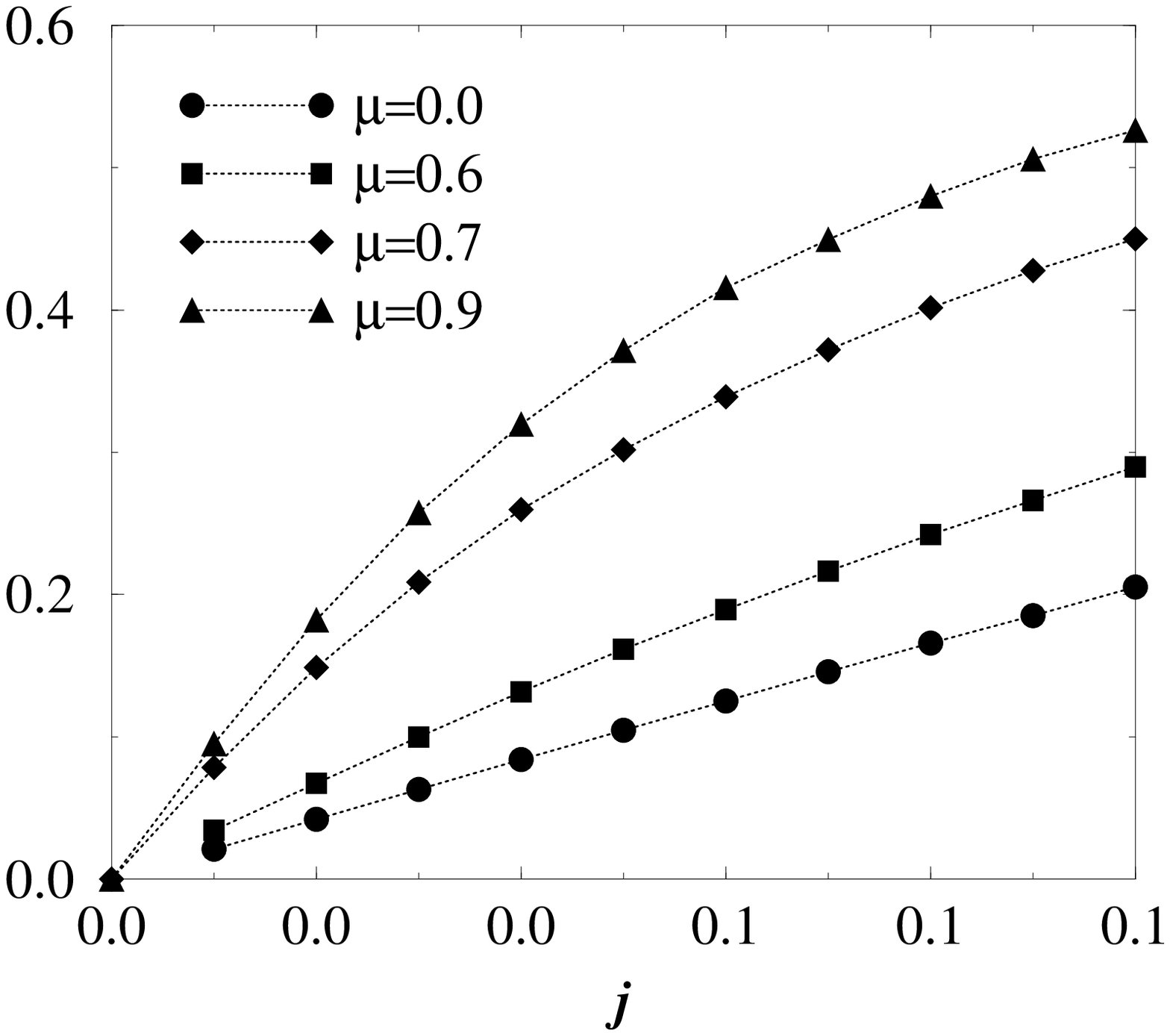}{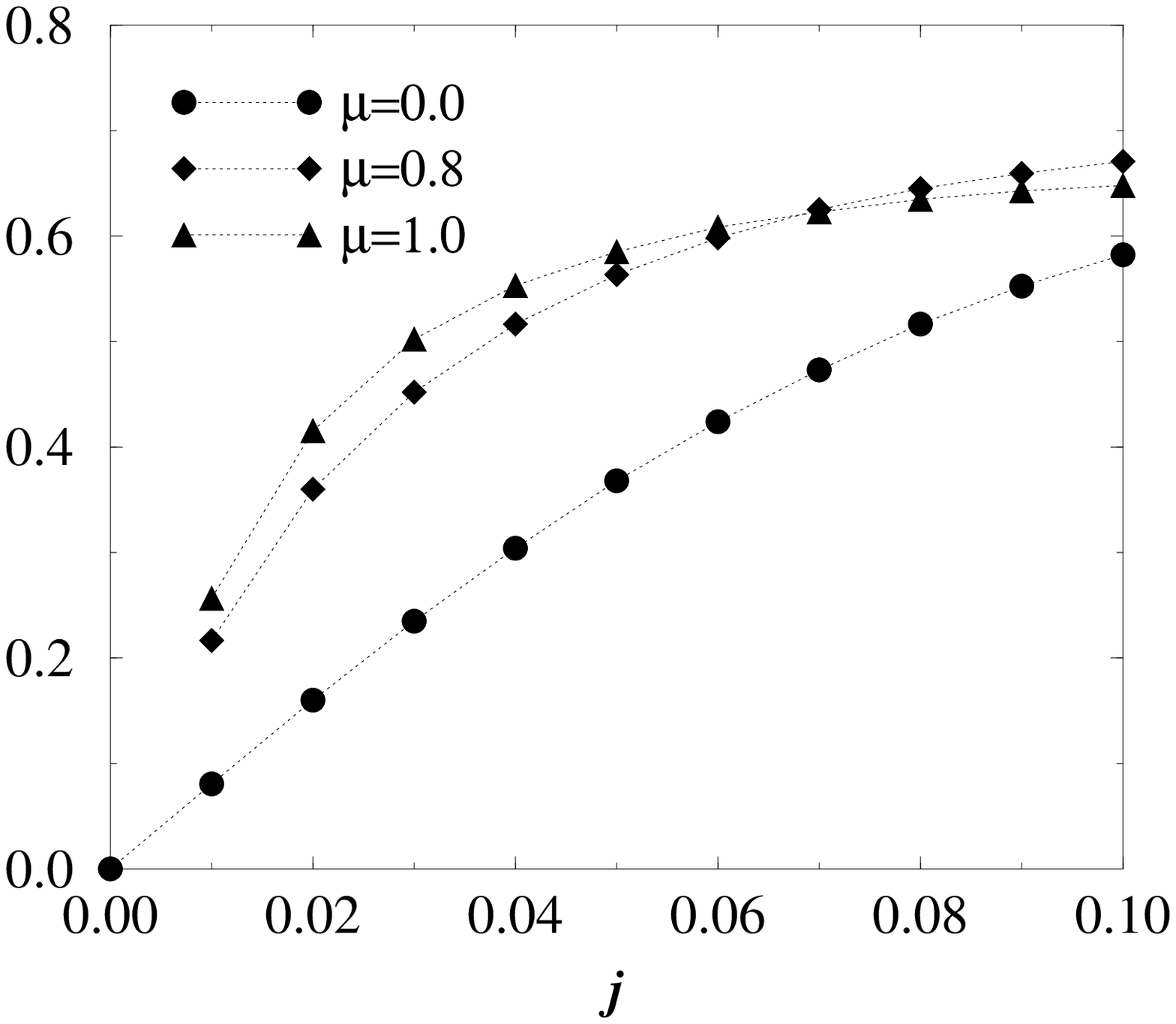}
\vskip -0.5cm
\caption{\hskip 4.1cm Figure 4:}
\vskip -0.5cm
\end{figure}

Whilst we have still 
not found unambiguous evidence for diquark condensation, it is
clear {\sl something\/} interesting is happening in the dense phase, as
revealed both by the long range timelike order in the behaviour of the two-point
function, and the non-linear behaviour of the one-point function with $j$.
Systematic effects due to the influence of Goldstone
modes, the relatively small number of states close to the Fermi surface on a
finite lattice in a small number of dimensions, and the quenched nature of the 
$\langle qq\rangle$ measurement may all need to be understood.
Interestingly, evidence for diquark condensation is much more compelling 
in simulations of SU(2) gauge theory \cite{MH}.
It is important, however, to understand how a BCS-like mechanism manifests
itself in simple four-fermion 
models before QCD can be tackled (recall that in this case
the condensate is not even gauge invariant). In future work we plan
to explore the spectroscopy of the model using the Gor'kov representation, 
and hopefully measure the gap, which is after all the quantity closest to 
physics.

\section*{Acknowledgments}
This work was supported by the TMR network ``Finite temperature phase
transitions in particle physics'' EU contract ERBFMRX-CT97-0122.

\end{document}